\documentclass[a4paper,twoside,12pt]{article}

\usepackage{amsmath,exscale,amsthm}
\usepackage{latexsym,fancyhdr}
\usepackage{graphicx,psfrag}
\usepackage[small,bf,hang]{caption}
\usepackage{cite}

\begin{document}

\setcounter{page}{1}
\pagenumbering{arabic}

\title{Some Heuristic Semiclassical Derivations of the Planck Length, the Hawking Effect
and the Unruh Effect}
\author{Fabio Scardigli}
\date{}
\maketitle
\begin{center}
\textit{Dipartimento di Fisica dell'Universita', \\ Via Celoria 16, 20133 Milano, Italy.}
\end{center}
\begin{abstract}
The formulae for Planck length, Hawking temperature and Unruh-Davies temperature are
derived by using only laws of classical physics together with the Heisenberg principle
put in the form $\Delta E \Delta x \simeq \hbar c/2$. Besides, it is shown how the Hawking 
relation can be deduced from the Unruh relation by means of the principle of equivalence;
the deep link between Hawking effect and Unruh effect is in this way clarified.    
\end{abstract}
\section{Introduction.}
In this short communication we shall expose several simple deductions of the fundamental 
relations regarding the Planck length, the temperature of the event horizon of a black hole
(Hawking effect) and the temperature measured by a detector placed in a uniformly accelerated 
frame in the flat space-time (Unruh effect). All these derivations will be obtained by 
employing simple relations of classical physics ~\cite{boyer1}, together with the elementary 
expressions of the Heisenberg uncertainty principle, of the relativistic energy-mass equivalence, 
and of the Boltzmann relation of the kinetic theory of gases, which links the 
kinetic energy of a particle to the absolute temperature.\\
The system of unities employed here is c.g.s.

\section{The Planck length.}    
In order to obtain this fundamental length we use the energy-time form of the Heisenberg 
principle: $\Delta E \Delta t \simeq \hbar/2$ ~\cite{Caldirola}. As it is well known, if the mean 
life of an excited state is $\Delta t$, then its energy is undetermined for a quantity 
$\Delta E$. We can put the principle in the form $\Delta E \Delta x \simeq \hbar c/2$; 
this will be useful in the deduction of the Planck length $L_P$.\\
$L_P$ represents the typical space length around which the quantum fluctuations 
become essential for the determination of the geometry of space-time 
~\cite{dewitt},~\cite{wheeler}. At this scale, strong distortions and rips appear in the 
space-time topology and (virtual) black holes can be formed ~\cite{kuchar}.\\
Suppose to observe a space region of width $\Delta x$; the energy fluctuations of the 
quantum vacuum in this region will be of the order of $\Delta E \simeq \hbar c/(2\Delta x)$.
This energy (i.e. the mass $\Delta M = \Delta E/c^2$) is confined in a region of width 
$\Delta x$. Besides, suppose that this length is less than (or of the order of) the 
Schwarzschild radius associated with the mass $\Delta M$: $R=(2G \Delta M)/c^2$. 
In this case the mass should be confined in the inner of its event horizon and a (virtual)
black hole should be formed. This length will be, therefore,
\begin{equation}
\Delta x=\frac{2G\Delta M}{c^2}=\frac{2G\Delta E}{c^4}=\frac{G \hbar}{c^3 \Delta x}
\end{equation}
or
\begin{equation}
\Delta x=\left(\frac{G \hbar}{c^3}\right)^{1/2}.
\end{equation}
As one may see, at this space lenth there are so strong energetic fluctuations of the 
quantum vacuum that virtual black holes are created, and these are distorsions and 
rips in the tissue of space-time. This is the so-called Planck length: 
$L_P=(G \hbar/c^3)^{1/2}$.
\section{The Hawking effect.}
Now we consider a Schwarzschild black hole with a radius $R=2GM/c^2$. Imagine a particle
(for example, an electron) with a mass $m$, near to the horizon. Its potential energy 
due to gravity is (classically)
\begin{equation}
U=\frac{GMm}{R}.
\end{equation}
On th external edge of the event horizon, this particle feels a potential gradient equal to
(for a small radial displacement $\Delta x$)
\begin{equation}
\Delta U=\frac{GMm}{R^2}\Delta x.
\end{equation}
The kinetic energy $\Delta E_c$ acquired by the particle in the shift $\Delta x$ is equal
to the lost potential energy $\Delta U$. We require that $\Delta E_c$ is of the order of that 
needed for the creation of a particle-antiparticle pair from the quantum vacuum, i.e. we put
\begin{equation}
\Delta E_c = 2mc^2.
\end{equation}
Therefore
\begin{equation}
\frac{GMm}{R^2}\Delta x = 2mc^2\quad \quad {\rm or} \quad\quad \Delta x=2c^2 \frac{R^2}{GM}.
\end{equation}           
This process, repeated many times, will create a gas of particles on the external edge of 
the event horizon. Since the particles are moving in a space slice of thickness $\Delta x$, 
each of them has an uncertainty in the kinetic energy equal to 
\begin{equation}
\Delta E = \frac{\hbar c}{2 \Delta x}=\frac{\hbar GM}{4cR^2}=\frac{\hbar c^3}{16 GM}.
\end{equation}           
Every particle, thought as a classical particle, has a mean kinetic energy of 
\begin{equation}
E_c =\frac{3}{2}kT,
\end{equation}           
where $T$ is the temperature of this gas of particles. If we interpret the uncertainty 
in the kinetic energy as due to the thermal agitation, we can write
\begin{equation}
\frac{3}{2}kT=\frac{\hbar c^3}{16 GM}
\end{equation}               
or, rather,
\begin{equation}
T=\frac{\hbar c^3}{24 k GM}.
\label{HRel}
\end{equation}               
This is the temperature of the particle gas (for example, electronic gas) thickened 
on the surface of the event horizon, that is the temperature of the hole ~\cite{sciama}
as seen by a quasi minkowkian observer set up at infinity. We remind that the exact 
quantum-mechanical calculus performed by Hawking provides the relation ~\cite{hawking}
\begin{equation}
T=\frac{\hbar c^3}{8 \pi k GM}.
\end{equation}               
The agreement, as one can see ($8 \pi\approx 24$), is very good.
\section{The Unruh effect.}
The Unruh-Davies effect \cite{Davies}, \cite{Unruh} consists in the fact 
that the response of a detector subjected to a uniform acceleration in a flat spacetime 
is the same as it would be if the detector were put in a thermal bath with a temperature 
of $T=\hbar a/2 \pi c k$, where $a$ is the acceleration of the detector ~\cite{sciama}.
In other words, an observer who undergoes a uniform acceleration $a$, apparently sees a 
fixed surface radiate with a temperature of $T=\hbar a/2 \pi c k$.\\
This relation can be derived in the following way.\\
Consider an Einstein lift uniformly accelerated towards the top. Suppose that near to 
the bottom floor of the lift there are several electrons, jointed with the lift.
The translatory kinetic energy acquired by one of these particles during the space shift 
$\Delta x$ is
\begin{equation}
\Delta E_c = mv \Delta v = ma \Delta x ,
\end{equation}
where $a$ is the acceleration of the lift, and, consequently, of the electrons.
We require that this energy is equal to that needed to create a pair $e^{+}e^{-}$
from the quantum vacuum; so we have to put 
\begin{equation}
\Delta E_c = 2mc^2 .
\end{equation}
From here we can obtain the amplitude of the shift $\Delta x$ along which the electron 
is to be accelerated in order to acquire such energy:
\begin{equation}
ma \Delta x = 2mc^2 \quad \quad \Longrightarrow \quad\quad \Delta x = \frac{2c^2}{a}.
\end{equation}
The pairs $e^{+}e^{-}$ created in this way are confined in a space slice of amplitude 
$\Delta x$. The uncertainty in the kinetic energy of the single electron is therefore 
\begin{equation}
\Delta E = \frac{\hbar c}{2\Delta x}=\frac{\hbar a}{4 c}.
\end{equation}
This energy is classically interpretable as kinetic energy due to the thermal agitation 
of the leptons. That is to say, for the single electron we can write
\begin{equation}
\frac{3}{2}kT=\frac{\hbar a}{4 c}.
\end{equation}
Hence
\begin{equation}
T=\frac{\hbar a}{6 k c},
\label{URel}
\end{equation}
where $T$ evidently represents the temperature of the electronic gas detected by 
the accelerated observer (see also on this subject \cite{boyer2}, ~\cite{boyer3}).\\
Still, we note that the relation here obtained is in a near perfect agreement with 
the relation worked out by Davies \cite{Davies} and Unruh \cite{Unruh}, i.e.
\begin{equation}
T=\frac{\hbar a}{2\pi k c}.
\end{equation}
In fact, $2 \pi \approx 6$.
\section{Applications of the principle of equivalence.}

The Unruh effect also takes place in a gravitational field, as one can see in an 
easy way by applying the principle of equivalence \cite{boyer4}. A direct way to 
deduce this result, is to consider the gravitational acceleration measured by an 
observer which stands in the gravitational field of a mass $M$: classically $g=GM/r^2$.
The potential energy acquired by a particle of mass $m$ after a radial shift
$\Delta x$ in the gravitational field is
\begin{equation}
\Delta U = gm \Delta x.
\end{equation}
This energy will be sufficient for the creation of a particle antiparticle pair 
(e.g. $e^{+}e^{-}$) if
\begin{equation}
gm \Delta x = 2mc^2,
\end{equation}
or
\begin{equation}
\Delta x=\frac{2c^2}{g}.
\end{equation}
As shown before, the uncertainty in energy of each one of these particles is
\begin{equation}
\Delta E = \frac{\hbar c}{2 \Delta x}=\frac{\hbar g}{4c}.
\end{equation}
This is again interpretable as thermal kinetic energy of the particle. Therefore,
the temperature of the particle gas, (created in this way), measured by the 
observer in the gravitational field $g$ is
\begin{equation}
T = \frac{\hbar g}{6ck}.
\end{equation}
Of course, the principle of equivalence ensures us that this is also the temperature
experimented by an observer jointed with a uniformly accelerated frame, with acceleration 
$\vec{a}=-\vec{g}$.\\
Besides, from the Unruh relation (\ref{URel}), it is possible to infer, via the equivalence 
principle, the Hawking relation (\ref{HRel}). In fact, an uniformly accelerated observer 
experiments a vacuum temperature of 
\begin{equation}
T = \frac{\hbar a}{6ck}.
\end{equation}
Therefore, for the equivalence principle, our observer, standing in a gravitational field, 
will measure a temperature of
\begin{equation}
T = \frac{\hbar g}{6ck},
\end{equation}
where $\vec{g}=-\vec{a}$.
If the reference frame is close to the event horizon of a Schwarzschild black hole,
the field strength has the value
\begin{equation}
g_{SCH}=\frac{GM}{R^2_{SCH}}=\frac{c^4}{4GM}
\end{equation}
and therefore the temperature of the horizon is 
\begin{equation}
T = \frac{\hbar g}{6ck}=\frac{\hbar c^3}{24 kGM},
\end{equation}
which is just the Hawking temperature of the black hole.\\
This fact stresses the link between Unruh effect and Hawking 
effect, and the latter appears clearly to be a particular case of the former.        

\begin{thebibliography}{99}
\bibitem{boyer1} Boyer T.H., \textit{Le Scienze}, \textbf{35}, (1985) 68. 
\bibitem{Caldirola} Caldirola P., Prosperi G.M., \textit{Introduzione alla Fisica Teorica}
(UTET, Torino) 1982, pp. 509-510.
\bibitem{dewitt} See, for example, De Witt B.S., \textit{Sci.Am.}, \textbf{249} (1983) 104.
\bibitem{wheeler} Wheeler J.A., \textit{Geometrodynamics} (Academic Press, New York, N.Y.) 1962. 
\bibitem{kuchar} Kuchar K., \textit{J. Math. Phys.}, \textbf{11} (1970) 3322.
\bibitem{sciama} Sciama D.W., Candelas P., Deutsch D., \textit{Adv. Phys.}, \textbf{30} (1981) 327.
\bibitem{hawking} Hawking S.W., \textit{Commun. Math. Phys.}, \textbf{43} (1975) 199.
\bibitem{Davies} Davies P.C., \textit{J. Phys. A}, \textbf{8} (1975) 609.
\bibitem{Unruh} Unruh W.G., \textit{Phys. Rev. D}, \textbf{14} (1976) 870.
\bibitem{boyer2} Boyer T.H., \textit{Phys. Rev. D}, \textbf{21} (1980) 2137. 
\bibitem{boyer3} Boyer T.H., \textit{Phys. Rev. D}, \textbf{29} (1984) 1089
\bibitem{boyer4} Boyer T.H., \textit{Phys. Rev. D}, \textbf{29} (1984) 1097.
\end{thebibliography}
\end{document}